\documentclass[reprint,aps,pra,superscriptaddress,
amsmath,amssymb,nobibnotes]{revtex4-1}
\usepackage{mathtools}
\usepackage{graphicx}
\usepackage{booktabs,multirow}
\usepackage{braket}
\usepackage{xcolor}
\usepackage{textcomp}

\usepackage[running]{lineno}
\graphicspath{ {figures_for_paper/} }
\usepackage[caption=false]{subfig}
\usepackage{braket}
\usepackage[
  bookmarks=false,
  colorlinks,
  linkcolor=blue,
  urlcolor=blue,
  citecolor=blue,
  plainpages=false,
  pdfpagelabels,
  final,
  breaklinks=true
]{hyperref}

\begin{document}
\title{Experimental demonstration of superresolution of partially coherent light sources using parity sorting}
\author{S. A. Wadood}
\affiliation{The Institute of Optics, University of Rochester, Rochester, New York 14627, USA}
\affiliation{Center for Coherence and Quantum Optics, University of
Rochester, Rochester, New York 14627, USA}
\author{Kevin Liang}
\affiliation{The Institute of Optics, University of Rochester, Rochester, New York 14627, USA}
\affiliation{Center for Coherence and Quantum Optics, University of
Rochester, Rochester, New York 14627, USA}
\author{Yiyu Zhou}
\affiliation{The Institute of Optics, University of Rochester, Rochester, New York 14627, USA}
\affiliation{Center for Coherence and Quantum Optics, University of
Rochester, Rochester, New York 14627, USA}
\author{Jing Yang}
\affiliation{Center for Coherence and Quantum Optics, University of
Rochester, Rochester, New York 14627, USA}
\affiliation{Department of Physics and Astronomy, University of Rochester, Rochester, New York 14627, USA}
\author{M. A. Alonso}
\affiliation{The Institute of Optics, University of Rochester, Rochester, New York 14627, USA}
\affiliation{Center for Coherence and Quantum Optics, University of
Rochester, Rochester, New York 14627, USA}
\affiliation{Aix Marseille Univ, CNRS, Centrale Marseille,
Institut Fresnel, UMR 7249, 13397 Marseille Cedex 20, France}
\author{X.-F. Qian}
\affiliation{Department of Physics and Center for Quantum Science and Engineering,
Stevens Institute of Technology, Hoboken, NJ 07030, USA}
\author{T. Malhotra}
\email{Currently with Facebook Reality Labs, Redmond, WA, USA}
\affiliation{Center for Coherence and Quantum Optics, University of
Rochester, Rochester, New York 14627, USA}
\affiliation{Department of Physics and Astronomy, University of Rochester, Rochester, New York 14627, USA}
\author{S. M. Hashemi Rafsanjani}
\affiliation{Department of Physics, University of Miami, Coral Gables, Florida 33146, USA}
\author{Andrew N. Jordan}
\affiliation{Center for Coherence and Quantum Optics, University of
Rochester, Rochester, New York 14627, USA}
\affiliation{Department of Physics and Astronomy, University of Rochester, Rochester, New York 14627, USA}
\affiliation{Institute for Quantum Studies, Chapman University, Orange, California 92866, USA}
\author{Robert W. Boyd}
\affiliation{The Institute of Optics, University of Rochester, Rochester, New York 14627, USA}
\affiliation{Center for Coherence and Quantum Optics, University of
Rochester, Rochester, New York 14627, USA}
\affiliation{Department of Physics and Astronomy, University of Rochester, Rochester, New York 14627, USA}
\affiliation{Department of Physics, University of Ottawa, Ottawa, Ontario K1N 6N5, Canada}
\author{A. N. Vamivakas}
\email{nick.vamivakas@rochester.edu}
\affiliation{The Institute of Optics, University of Rochester, Rochester, New York 14627, USA}
\affiliation{Center for Coherence and Quantum Optics, University of
Rochester, Rochester, New York 14627, USA}
\affiliation{Department of Physics and Astronomy, University of Rochester, Rochester, New York 14627, USA}
\affiliation{Materials Science, University of Rochester, Rochester, NY 14627, USA}

\date{\today}

\begin{abstract}
Analyses based on quantum metrology have shown that the ability to localize the positions of two incoherent point sources can be significantly enhanced through the use of mode sorting. Here we theoretically and experimentally investigate the effect of partial coherence on the sub-diffraction limit localization of two sources based on parity sorting. With the prior information of a negative and real-valued degree of coherence, higher Fisher information is obtained than that for the incoherent case. Our results pave the way to clarifying the role of coherence in quantum limited metrology.
\end{abstract}



\maketitle
\section{Introduction}
The resolution of imaging systems is limited by the size of the diffraction-limited point spread function (PSF) \cite{goodman_Fourier_Optics_book}. To quantify this resolution, the Rayleigh criterion has been proposed and widely used \cite{rayleigh1879xxxi}. Recently, the analysis of optical resolution has been recast in terms of Fisher Information (FI) \cite{tsang2016quantumtheoryofsuperresolution,zhou2019modern,yang2019optimal}, which quantifies the precision of measurements and is inversely proportional to the parameter estimation error. Generally, the FI of the estimation of separation $\delta$ between two spatially incoherent point sources depends on the type of measurement performed on the image plane field. In the case of direct detection of image plane intensity, the FI goes to zero as $\delta \xrightarrow {}0$, an effect termed as Rayleigh's curse. In their seminal work \cite{tsang2016quantumtheoryofsuperresolution}, Tsang et al. showed that Rayleigh's curse can be overcome if the optical field is detected by an appropriate spatial mode demultiplexer (SPADE), given prior knowledge of two equally bright and incoherent point sources versus a single emitter. The FI for such a scheme is constant as $\delta \xrightarrow{}0$, as has been verified experimentally\cite{sanchezsoto_2016_optica,Steinberg2017BeatingRayleighscurse,superresolution_with_heterodyne,OptExpress_Parity_Sorting2016,yiyu2019axial_superresolution_optica}.\par
\begin{figure*}[t]
\centering
\includegraphics[scale=0.80]{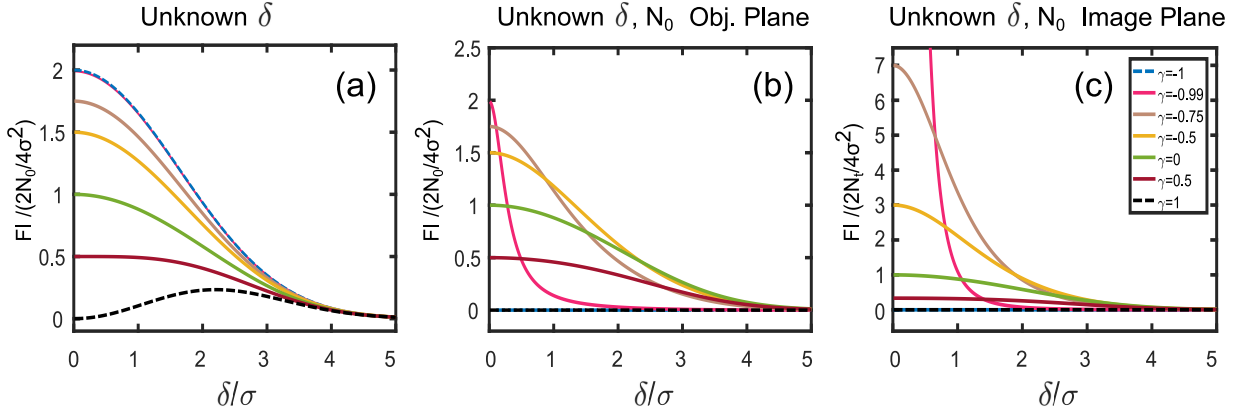}
\caption{Expected FI for Parity Sorter plotted versus $\delta/\sigma$. A higher FI corresponds to a lower estimation error. a: FI prediction for the case when $\delta$ is the only unknown parameter. For this case, $\gamma=-1$ gives the highest FI (dashed blue line on top of the $\gamma=-0.99$ curve), as predicted by the Tsang--Nair model \cite{tsangcommentonSaleh}. b: FI prediction for the case of unknown input photon number $N_{0}$. For this case, the FI is zero for $|\gamma|=1$. As $\gamma\xrightarrow[]{}-1$, the FI curve gets concentrated near $\delta=0$, but is still bounded above by twice the FI for $\gamma=0$. The curves in (a,b) are normalized by the object plane photon numbers. c: FI prediction normalized by the image plane photon number for the case of unknown $N_{0}$. These curves are related to the curves in (b) by the weight factor of $(1+\gamma d)$ as explained in the text. As $\gamma\xrightarrow[]{}-1$, this image plane FI diverges and gets concentrated around $\delta=0$, a result which was predicted using a quantum calculation in \cite{SalehreplytoTsang,SanchezSoto_fisherinformation_With_Coherence_Optica}. As explained in the text, the information conveyed by curves (b,c) is the same. Note that the $\gamma=0$ curves (green line) are same in all the figures.}
\label{fig::1_partial_coherence_paper}
\end{figure*}
The sources, however, can have a non-zero coherence between them \cite{MandelandWolf}.  In fact, spatial coherence is a key parameter affecting the resolution of imaging systems \cite{goodman2015statisticaloptics}; coherent illumination techniques can offer enhanced resolution in microscopy \cite{CoherentMicroscopyBook} and two-point direct imaging \cite{BrianThompson_2_point_resolution_with_partially_coherent_light,nayyar_and_varma_1978twopointresolution}. Moreover, coherence imaging can offer significant practical advantages over conventional direct imaging systems, for example in the very long baseline radio interferometry (VLBI) used for black hole imaging \cite{blackhole_paper_2019}. It is then natural to ask how spatial coherence between the two sources affects the resolution obtained by SPADE. Recent theoretical works have extended the scope of the two-point estimation problem to include the general case of partial coherence among the two sources \cite{SalehResurgencePaper,tsangcommentonSaleh,SalehreplytoTsang,lee2019_SPIE_surpassing_FI_for_PartialCoherence,SanchezSoto_fisherinformation_With_Coherence_Optica,Kevin_multiparameter_Estimation}. In particular, it was shown that Rayleigh's curse can still be avoided for a known degree of spatial coherence $\gamma$ \cite{tsangcommentonSaleh,SalehreplytoTsang,Kevin_multiparameter_Estimation}. For the case of $\gamma<0$, an even greater sensitivity for SPADE was predicted than the incoherent case. The increased sensitivity needs to be carefully interpreted, taking into account photon budgeting considerations \cite{SanchezSoto_fisherinformation_With_Coherence_Optica}. Experimental demonstration of SPADE with partial coherence, however, has been lacking. The main result of our work is to experimentally demonstrate the breaking of Rayleigh's curse for partially coherent light sources using SPADE. In doing so, we also distill and connect the different elements of previous theoretical works.\par 
In Section \ref{sec:theory}, we derive the classical FI of our experimental setup for partially coherent fields. Special attention is paid to a priori assumptions and how they affect the obtained FI. The connection between previous works is also made clear in this section.
Section \ref{sec::Experimental Section} explains the experimental setup, the generation of spatial coherence, and a discussion of estimation statistics.
Section \ref{sec::Conclusion} summarizes the results. 
\par
\section{Theory}\label{sec:theory}
In this section we outline the calculation of the classical FI for parity sorting of the partially coherent field. Note that parity sorting falls under the scheme of binary SPADE (BSPADE), which is a family of measurements that simplifies SPADE at the cost of losing large-delta ($\delta>\sigma$) information \cite{tsang2016quantumtheoryofsuperresolution,FIO_2018_jeremyhassett2018}. For $\delta\ll\sigma$, it has been shown that a measurement of the even and odd projections of the input field has an FI that converges to the quantum optimal FI \cite{Steinberg2017BeatingRayleighscurse,OptExpress_Parity_Sorting2016}. We show explicitly how different a priori assumptions yield different FI curves. The physical problem is the following: Two point sources separated by $\delta$ and having a degree of spatial coherence $\gamma$ are imaged by an imaging system with a finite-sized aperture. The goal is to perform quantum-limited estimation of $\delta$ in the sub-Rayleigh regime by performing parity sorting on the image plane field. \par 
A partially coherent field is described by its cross-spectral density (CSD) function $W(x_{1},x_{2})$ \cite{MandelandWolf}. To proceed, we first note that $W$ can be decomposed via the coherent mode decomposition (CMD) \cite{wolf1981CMD}. For our problem, the simplest choice of modes is to decompose the $W$ in the symmetric (in phase) and antisymmetric (out of phase) combinations of the two sources. In the image plane, $W(x_{1},x_{2})$ is given as 
\begin{linenomath*}
\begin{align}\label{eq:MutualCoherence}
    W(x_{1},x_{2})=N_{0}\kappa\sum_{k=1}^2 p_{k}\phi^{*}_{k}(x_{1})\phi_{k}(x_{2}),
\end{align}
\end{linenomath*}
where $N_{0}$ is the average image plane photon number emitted by each point source, $\kappa$ is a space-invariant efficiency factor dictated by the aperture loss, $\phi_{k}(x)=f_{+}(x)-e^{i k \pi}f_{-}(x)$ are the symmetric ($k=1$) and antisymmetric ($k=2$) coherent modes, $f_{\pm}(x)=f(x\pm\delta/2)$ are the two point spread functions separated by $\delta$ - the parameter to be estimated, $p_{k}$ is a real number such that $0\leq p_{k}\leq 1$, and $p_{1}+p_{2}=1$. In what follows, the terms even and odd modes are used interchangeably with symmetric and antisymmetric modes. We assume Gaussian PSFs of width $\sigma$ such that $f(x)=\frac{e^{-x^2/4\sigma^2}}{(2\pi\sigma^2)^{1/4}}$ is the field PSF.
The total number of photons in the image plane is given by
\begin{linenomath*}
\begin{align}\label{eq:total_photon_number}
    N_{t}=\int dx W(x,x)=2N_{0}\kappa(1+\gamma d),
\end{align} 
\end{linenomath*}
where  $d=\int dx f_{+}(x)f_{-}(x)=e^{-\delta^2/(8\sigma^2)}$ is the overlap integral of the two shifted PSFs, and $\gamma=p_{1}-p_{2}$ is an effective degree of spatial coherence between the two sources. It is here that we first encounter the departure from the incoherent estimation problem; for $\gamma\neq0$, $N_{t}$ depends on the parameter $\delta$ to be estimated. Hence, it is necessary to spend some time clarifying the interpretation of the FI for partially coherent sources. For a parity sorter, the photon numbers in the even and odd ports are, respectively,
\begin{linenomath*}
\begin{equation}\label{eq:modal_weights}
\begin{aligned}
    N_{1}&=N_{0}\kappa p_{1}\int dx  |\phi_{1}(x)|^2=N_{0}\kappa(1+\gamma)(1+d),\\
    N_{2}&=N_{0}\kappa p_{2}\int dx  |\phi_{2}(x)|^2=N_{0}\kappa(1-\gamma)(1-d).
\end{aligned}
\end{equation}
\end{linenomath*}
Equations (\ref{eq:total_photon_number},\ref{eq:modal_weights}) are derived in the supplement. We assume that $\gamma,\kappa$ are known a priori. If we know $N_{0}$ and the only unknown in the experiment is $\delta$, then assuming Poisson statistics it can be shown \cite{tsang2016quantumtheoryofsuperresolution} that the FI for parity sorting is given by
\begin{linenomath*}
\begin{align}\label{eq:FI_Tsang}
    \frac{F_{\delta}(\delta,\gamma)}{2N_{0}\kappa}=\frac{\delta^2 d^2}{16\sigma^4}\left(\frac{1-\gamma d}{1-d^2}\right),
\end{align}
\end{linenomath*}
where the subscript $\delta$ denotes that $\delta$ is the unknown parameter. Note that $F_{\delta}(\delta,\gamma)$ is normalized by $2N_{0}\kappa$, the total object plane photons multiplied by the loss factor. $F_{\delta}(\delta,\gamma)$ is plotted in Fig. (\ref{fig::1_partial_coherence_paper}a), and $\kappa$ has been absorbed into $N_{0}$ for the plot. These curves show that the highest FI is achieved for $\gamma=-1$. The physical operation of parity sorting affords some intuition about this FI behavior. For $\gamma=-1$, all photons are routed to the odd port, and we have $N_{1}=0$ and $N_{2}=2N_{0}\kappa(1-d)$. Knowing the total emitted photon number $2N_{0}$ and the total detected photon number $N_{2}$ allows us to estimate $\delta$ directly. For $\delta\ll\sigma$, the power in the odd port is well approximated as $N_{0}\kappa(1-\gamma)\delta^2/8\sigma^2$. Thus for sub-Rayleigh separation, the odd port has the most photons for $\gamma=-1$, and hence the highest FI.\par
It is not uncommon, however, that an experimentalist only has access to image plane photons, and does not have knowledge of $N_{0}$. When both $\delta$ and $N_{0}$ are unknown, the FI is found from the multiparameter Cramer--Rao bound (CRB); this FI is given by
\begin{linenomath*}
\begin{align}\label{eq:FI_Object_plane}
    \frac{{F}_{\delta,N_{0}}(\delta)}{2N_{0}\kappa}=\frac{\delta^2d^2}{16\sigma^4(1+\gamma d)}\left(\frac{1-\gamma^2}{1-d^2}\right),
\end{align}
\end{linenomath*}
and is plotted in Fig. (\ref{fig::1_partial_coherence_paper}b). Note that as $\gamma\xrightarrow{}-1$, $F_{\delta,N_{0}}(\delta,\gamma)$ becomes concentrated near $\delta=0$. While Rayleigh's curse is avoided for $\gamma<0$, i.e., $F_{\delta,N_{0}}(\delta=0,\gamma)=(1-\gamma)/4\sigma^2$, the FI is effectively zero for all $\delta\neq0$ and $\gamma=-1$. Figures (\ref{fig::1_partial_coherence_paper}a,b) clearly show how the knowledge or ignorance of the object plane photon number affects the FI for $\delta$ estimation in the presence of partial coherence.\par
We can now ask the more practical question of how to estimate $\delta$ when we only detect the image plane field, and have no knowledge of $N_{0}$ ? In this case one can use the normalized modal weights $p_{1,2}=N_{1,2}/N_{t}$ which are \textit{independent of $N_{0}$}. The statistics are described in this case by a binomial likelihood function \cite{tsang2016_SPIE_quantum_information_for_semiclassical_optics}. We can then calculate the \textit{image plane} FI by the formula
\begin{linenomath*}
\begin{align}\label{eq:FI_Image_plane}
    F_{img}(\delta,\gamma)&=\sum_{i=1,2}\frac{1}{p_{i}(\delta,\gamma)}\bigg(\frac{\partial p_{i}(\delta,\gamma)}{\partial \delta}\bigg)^2\nonumber\\
    &=\frac{\delta^2d^2}{16\sigma^4(1+\gamma d)^2}\left(\frac{1-\gamma^2}{1-d^2}\right),
\end{align}
\end{linenomath*}
where the subscript `img' denotes image plane and the function is plotted in Fig. (\ref{fig::1_partial_coherence_paper}c). We emphasize that $F_{img}$ is normalized per image plane photon; physically, Eq. (\ref{eq:FI_Image_plane}) quantifies the information provided by a single photon in the image plane, and is agnostic to the number of object plane photons. Figure (\ref{fig::1_partial_coherence_paper}c) then shows that given equal number of photons in the image plane, $\gamma<0$ can offer increased sensitivity in the regions $\delta\ll\sigma$. Note that Eq. (\ref{eq:FI_Image_plane}) is related to Eq. (\ref{eq:FI_Object_plane}) by a simple `weight' factor of $(1+\gamma d)$, which also relates the image and object plane photon number in Eq. (\ref{eq:total_photon_number}). While the image plane FI might increase for $\gamma<0$, more object plane photons are needed to maintain a constant image plane photon number, a `cost' that is captured by the factor of $(1+\gamma d)$. The image plane FI is also zero for $\gamma=-1$, in which case all clicks occur at the odd port for all $\delta$. If the experimentalist does not know $N_{0}$, they do not get any information about $\delta$ from just measuring clicks at the odd port. In any case, Figs. (\ref{fig::1_partial_coherence_paper}b) and (\ref{fig::1_partial_coherence_paper}c) give the same information, as there is a one-to-one correspondence between the two curves. Alternatively, the lowerbound on the variance of an unbiased estimator can equivalently be found either from Eq. (\ref{eq:FI_Object_plane}) or Eq. (\ref{eq:FI_Image_plane}).\par
 Incidentally, the aforementioned discussion provides clarity to the debate between, among others, the Tsang--Nair (TN) model \cite{tsangcommentonSaleh} and the Larson--Saleh (LS) model \cite{SalehreplytoTsang}. Strictly speaking, the TN model assumes knowledge of $N_{0}$, while the LS model assumes an unknown $N_{0}$. Specifically, Fig. (\ref{fig::1_partial_coherence_paper}a) agrees with the TN model, and Fig. (\ref{fig::1_partial_coherence_paper}c) agrees with the LS model. Figure (\ref{fig::1_partial_coherence_paper}b) bridges the TN and LS models. We note that Hradil et. al. \cite{SanchezSoto_fisherinformation_With_Coherence_Optica} also advocated the use of the weighted version of image plane FI to take into account the image plane photon number variation with $\gamma,\delta$, and their results also imply the curves in Fig. (\ref{fig::1_partial_coherence_paper}b). Depending on the a priori assumptions afforded by the experimental setup, either TN or LS models will correctly describe the estimation statistics. Note that a similar observation has been made for coherent microscopy \cite{LauraWaller2016standardizingresolutionclaimsforcoherentmicroscopy}, which advocates the `mandatory inclusion of information about underlying \textit{a priori} assumptions' when discussing resolution claims.\par
Having clarified the issue of the FI interpretation for partial coherence, we can now proceed to discuss the experiment.  Realistically, we will use the image plane model as it reflects a common situation in imaging, microscopy, and astronomy. Note that realistic situations have more than just $\delta$ and $N_{0}$ as possible unknowns. For example, our analysis till now has assumed the presence of only two sources, equal intensities of the two sources, a known centroid of the objects to which the parity sorter is aligned, and, most importantly, a known $\gamma$. In practice, one needs a combination of direct imaging, coherence interferometry, and parity sorting to estimate these unknown parameters. The application of quantum metrology-inspired ideas such as SPADE to practical situations is an active field of research \cite{de2021discrimination_under_misalignment,PRL2020_superresolution_limits_frm_crosstalk,Fabre2020_Optica_SPADE_2D_Crosstalk,Michael_Grace_JOSAB_unknown_Centroid}. These considerations, however, are not relevant to our proof-of-principle experiment in which we consider only $\delta$ and $N_{0}$ as the unknown parameters.
\section{Experiment}\label{sec::Experimental Section}

\subsection{Spatial Coherence Generation}\label{sec::CMD Explanation Section}
\begin{figure}[t]
\centering
\hspace*{0cm} 
\includegraphics[trim={0cm 0cm 0cm 0cm},scale=1]{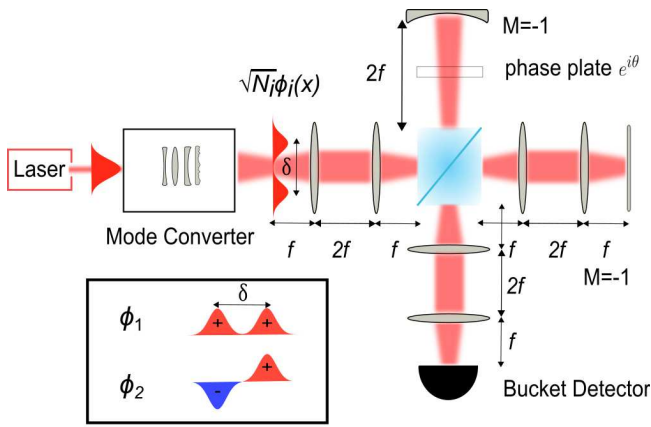}
\caption{Experimental Setup: A 795nm Gaussian beam is
converted to either a symmetric ($\phi_{1}$) or antisymmetric ($\phi_{2}$) mode, shown in the inset, via a mode converter consisting of linear optical transformations. The mode amplitudes are set to $\sqrt{N_{i}}=\sqrt{N_{t}p_{i}}$, with $p_{i}$ being the normalized modal weights and $N_{t}$ being an arbitrary image plane photon number, to generate the CSD given by Eq. (\ref{eq:MutualCoherence}). Polarization optics and attenuators, not shown in the figure, are used to control the power of the beam. At any given time, one of the coherent modes is sent to a Michelson type image inversion interferometer, which separates the even and odd components of the input field. One arm has a $4f$ system, which acts as an identity operator after the beam double passes it. The other arm has a $2f$ system, implemented by a convex mirror, and an extra quadratic phase, not shown, to cancel the defocus due to diffraction. This arm implements the transformation $(x,y) \xrightarrow{} (-x,-y)$. The combined beams from both arms are imaged onto a bucket detector. The power in the even and odd modes can be measured by setting the phase difference $\theta$ to 0 and $\pi$ respectively. In the experiment, all modes used are symmetric about the y axis such that $E(x,-y)=E(x,y)$. The interferometer then works as a parity sorter in the $x$-direction.
}
\label{fig::2_partial_coherence_paper}
\end{figure}
We use a parity sorter to perform SPADE on two spatially partially coherent sources. To generate partial coherence, we use the CMD \cite{wolf1981CMD}. Physically, such a CMD means that the spatial coherence at the input plane to the SPADE setup can be engineered by incoherently mixing appropriately scaled symmetric and antisymmetric modes. This can be realized by adding a path difference between coherent modes that is larger than the laser coherence length. Alternatively, we can `switch' between the modes in time, with the switching time longer than the laser coherence time, and add the recorded intensities digitally \cite{rodenburg2014CMD,RafsanjaniOL_CMD_Bessel_correlation}. The CMD therefore allows us to generate spatial coherence `offline', by performing the intensity summation electronically. To generate an intensity distribution corresponding to a specific $\gamma$ in Eqs. (\ref{eq:modal_weights}), we can post-select from a set of recorded intensities of $\phi_{1,2}$ modes. This allows a great simplification of the experiment with respect to the precise control of $\gamma$. Note that we are not changing the temporal coherence properties; all the beams used are quasimonochromatic and therefore temporally coherent.
\begin{figure*}[t]
\centering
\hspace*{-0.28cm} 
\includegraphics[trim={0cm 0cm 0cm 0cm},clip,scale=1]{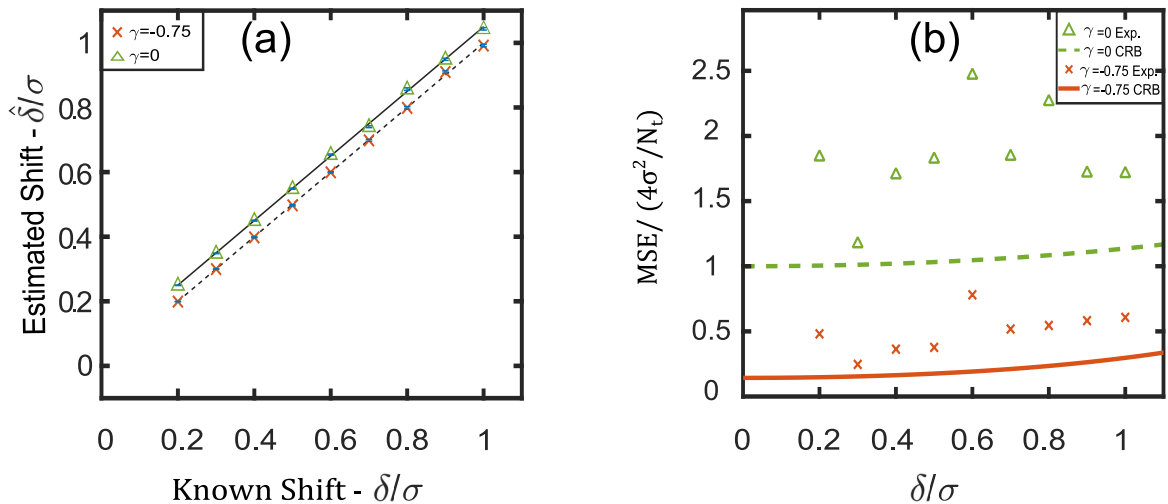}
\caption{a: Estimated shift $\hat{\delta}/\sigma$ for $\gamma=0,-0.75$ using MLE on the measured modal weights. The estimated shifts are all below the Rayleigh limit ($\delta=\sigma$). Each point represents the mean MLE of 100 measurements. The error bars are too small to be noticed on the graph, but are still bounded by the CRB as shown in (b). Note that the $\gamma=0$ estimates are not biased; to distinguish the two data sets, we introduce a vertical offset between the $\gamma=0$ and the $\gamma=-0.75$ estimates. Both $\gamma=0$ and $\gamma=-0.75$ estimates are in good agreement with the expected shifts. b: Measured MSE for $\gamma=0$ (green triangles) and $\gamma=-0.75$ (red crosses). For each data point, ML estimates from 100 trials were used to calculate the variance. Note that for a given $\delta/\sigma$, the MSE for $\gamma=-0.75$ is consistently less than the MSE for $\gamma=0$. The dashed green and solid red lines indicate the CRB for $\gamma=0,-0.75$ respectively. The CRB is given by the inverse of Eq. (\ref{eq:FI_Image_plane}). Technical noise factors causing the discrepancy between theory and experiment are explained in the main text.
}
\label{fig::3_partial_coherence_paper}
\end{figure*}
\subsection{SPADE using parity sorting}
\par After generating partially coherent fields, the next step is to perform parity sorting on the field described by Eq. (\ref{eq:MutualCoherence}). The experimental setup consists of an image inversion interferometer that sorts the input field based on its parity, as shown in Fig. (\ref{fig::2_partial_coherence_paper}). A Gaussian beam with $\sigma=327\pm4\mu m$ is converted into either a symmetric or antisymmetric mode using linear optics, which includes a spatial light modulator. The beam flux can be adjusted using polarization optics. The mode is presented to a Michelson type interferometer. The top arm, which includes a $2f$ imaging system and an extra quadratic phase implemented to cancel the defocus due to diffraction, implements the transformation $(x) \xrightarrow{} (-x)$ and the arm with the $4f$ system images the field with unity magnification, after two reflections. Experimental details of the interferometer are described in Ref. (\citenum{TanyaOpticsExpress}). For parity sorting, we set $\alpha=\pi$ in Eqs. (1-3) of Ref. (\citenum{TanyaOpticsExpress}). The field at the output of the interferometer is
\begin{linenomath*}
\begin{align}\label{eq:binarySPADE}
    E_{out}(x)=\frac{\sqrt{N_{k}}}{2}(\phi_{k}(x)+e^{i\theta}\phi_{k}(-x)),
\end{align}
\end{linenomath*}
where $k=1,2$, $\theta$ is the global phase difference between the two arms of the interferometer, $N_{k}$ is the photon number in the input mode $\phi_{k}$, and each $\phi_{k}$ is spatially coherent in Eq. (\ref{eq:binarySPADE}). 
Note that the coherent modes used are symmetric in $y$, so the 1D analysis is valid for the experiment. 

To project onto the even and odd components of the field, we can choose $\theta=0,\pi$. As explained in Section \ref{sec::CMD Explanation Section}, we send only one of the coherent modes $\phi_{k}$ at a given time. To generate CSD for a given $\gamma$, we add the measured intensities offline. Details of the offline coherence generation are given in the supplement. For $\theta=0 (\pi)$, all of the symmetric (antisymmetric) mode power will be directed to the bucket detector, while the antisymmetric (symmetric) mode will destructively interfere at the detector. For $\theta=0 (\pi)$ the output is called as the even (odd) port. A bucket detector measures the photon number in each port. 
\par
\subsection{Estimation Statistics}\label{sec::results}
The goal of superresolution is to estimate $\delta$ for regions of $\delta<\sigma$. To estimate $\delta$, we use maximum likelihood estimation (MLE) on the measured normalized modal weights $p_{1,2}$. Because we normalize the modal weights by the image plane photons, we use a binomial likelihood function for the parity sorter \cite{tsang2016_SPIE_quantum_information_for_semiclassical_optics}. The estimated $\hat{\delta}$ is shown in Fig. (\ref{fig::3_partial_coherence_paper}a). Note that all the estimated $\delta$'s are below the Rayleigh limit ($\delta=\sigma$). For $\delta$ in the interval $[0.2-1]\sigma$ (in increments of $0.1\sigma$), we take 100 images each of the symmetric and antisymmetric modes, thus getting 100 ML estimates and the corresponding variance.  We have not observed any bias in the estimates, as evident in Fig. (\ref{fig::3_partial_coherence_paper}a), where the mean of the estimates are equal to the true value of $\delta/\sigma$.
The variance in the MLE estimates, which is related the inverse of the FI, is too small to be noticed in Fig. (\ref{fig::3_partial_coherence_paper}a). Nevertheless, the variance of an unbiased estimator is lowerbound by the Cramer--Rao bound (CRB), which is related to the inverse of the FI.  Formally, $\textrm{Var}[\hat{\delta}]\geq (N_{t}F_{img})^{-1}$, where $\textrm{Var}[\hat{\delta}]$ is the variance in the MLE estimator $\hat{\delta}$, and $F_{img}$ is the image plane FI as given by Eq. (\ref{eq:FI_Image_plane}) and shown in Fig. (\ref{fig::1_partial_coherence_paper}c).  Figure (\ref{fig::3_partial_coherence_paper}b) shows the normalized Mean Square Error (MSE) $=N_{t}\textrm{Var}[\hat{\delta}]$ as a function of $\delta$ and two values of $\gamma=0,-0.75$. More importantly, Fig. (\ref{fig::3_partial_coherence_paper}b) shows that the MSE for $\gamma=-0.75$ is \textit{below} the CRB for the $\gamma=0$ case. In other words, not only is Rayleigh's curse avoided for $\gamma=-0.75$, the estimation is more precise than the incoherent case of $\gamma=0$. 
\if
REMOVE\{\{The mean number of photons used for each iteration is at least $10^5$. We can show explicitly that the increased sensitivity for the $\gamma=-0.75$ case is not due to a higher number of photons compared to the $\gamma=0$ case. To this end, the number of photons we use for each trial of the $\gamma=-0.75$ case is always \textit{less} than the number used for $\gamma=0$ case.\}\}

Fig. (\ref{fig::5_partial_coherence_paper}b) shows the measured FI, which is given by $(\textrm{MSE})^{-1}$.
\fi 
Note that the MSE are still offset from the CRB. To truly saturate the CRB, the system must be shot noise limited, and any other noise source will raise the MSE. Another source of noise in our system are the phase fluctuations in the interferometer when it is biased at $\theta=0~\textrm{or}~\pi$ (See Fig. (\ref{fig::2_partial_coherence_paper})). Furthermore, the MSE for $\gamma=0,-0.75$ might appear correlated, for example at $\delta=0.2,0.3$. This is because the same set of images are used for CMD of both $\gamma=0,-0.75$, and hence both $\gamma=0,-0.75$ MSE's will be affected by the same phase fluctuations; if the $\gamma=0$ MSE is higher, so will be the $\gamma=-0.75$ MSE. Finally, the CRB curves in Fig. (\ref{fig::3_partial_coherence_paper}b) are nearly equivalent to the quantum CRB predicted for $\delta<\sigma$ \cite{SalehreplytoTsang}, and therefore our measurements represent near quantum-limited localization of partially coherent sources.
\par
The reader might observe that no statistics for $|\gamma|=1$ are shown in Fig. (\ref{fig::3_partial_coherence_paper}). As discussed in Section (\ref{sec:theory}), the FI for $|\gamma|=1$ is zero for all $\delta$ if $N_{0}$ is unknown. The likelihood function in this case is independent of $\delta$ for $|\gamma|=1$, and hence $\delta$ cannot be estimated \textit{in principle}.$~N_{0}$ is unknown in our experiment because we generate the image plane field directly through unitary transformations and not through a Gaussian aperture that scales the coherent modes according to the $(1\pm d)$ factor in Eqs. (\ref{eq:modal_weights}). While our system has an effective `aperture' loss factor that connects the source photon number to the image plane photon number, this loss factor is independent of $\delta$ for the coherent modes generated by the SLM, as also reported in the Supplement. The experiment is the generalization of previous localization experiments on incoherent beams \cite{OptExpress_Parity_Sorting2016,Steinberg2017BeatingRayleighscurse}. This technique allows 1) a great experimental simplification with regard to avoiding the need to perform precise fabrication of point sources with different separations and 2) to circumvent issues of low photon budget and spurious diffraction effects from the source geometries. However, this technique fails to provide access to an effective object plane photon number which is related to the image plane photon number by the factor of $(1+\gamma d)$, and hence does not allow us to reconstruct results of Fig. (\ref{fig::1_partial_coherence_paper}a). Barring these technical difficulties, our theoretical and experimental results are easily generalized to the case of a known $N_{0}$. We note that having access to only the image plane photon number is however a common situation in optical physics, where one does not have an independent probe on the object plane photons. Our results are therefore valid for a large variety of microscopy and imaging experiments. The details of image processing, CMD, the photon number in Fig. (\ref{fig::3_partial_coherence_paper}) versus $\delta$, and mode generation are given in the Supplement.  
 \if 
 Therefore, our results show that for an unknown $\delta$:
\begin{itemize}
    \item For $|\gamma|=1$, the FI $\xrightarrow{}0$ as $\delta\xrightarrow{}0$.
    \item For $\gamma<0$, the FI is higher than the FI for $\gamma=0$. 
\end{itemize}
\fi
\section{Conclusion}\label{sec::Conclusion}
We have carried out a theoretical analysis of superresolution of partially coherent light using parity sorting. For partially coherent sources, the object plane photon number was identified as a relevant parameter that affects the obtainable FI, and that connects the different results of previous works \cite{tsangcommentonSaleh,SalehreplytoTsang,SanchezSoto_fisherinformation_With_Coherence_Optica}. We also performed parity sorting on two Gaussian PSFs with varying degrees of spatial coherence. Our results show that partial anticorrelation of the two sources increases the FI of $\delta$ estimation. Therefore, Rayleigh's curse can be avoided for partially coherent sources. The proof-of-principle experiment paves the way to using coherence as a resource in quantum-limited metrology. 
\if
To perform SPADE measurements, the normalization used in \cite{tsangcommentonSaleh} requires prior knowledge of $\delta$ - the parameter to be estimated.
\fi 
Our analysis assumes a real, known value of $\gamma$. Further studies could include concurrent estimation of $\delta$ and $\gamma$, for which a vanishing FI with $\delta\xrightarrow{}0$ is predicted \cite{SalehResurgencePaper,Kevin_multiparameter_Estimation}. The natural extension of the current work is to consider the more realistic case of multiparameter estimation of a complex $\gamma$, the centroid and intensity ratio of the two sources \cite{Kevin_multiparameter_Estimation}, and the effects of cross-talk in the SPADE setup \cite{Fabre2020_Optica_SPADE_2D_Crosstalk,PRL2020_superresolution_limits_frm_crosstalk,optimal_observables_for_practical_superresolution,de2021discrimination_under_misalignment}. While we have been primarily concerned with the two-point problem, the technique of SPADE can also tackle the more general problem of imaging an extended object scene. There the problem reduces to estimation of moments of the object in the sub-diffraction limit, a case which was treated for incoherent objects \cite{tsang2017subdiffraction,tsang2019quantum,zhou2019modern,tsang2020semiparametric}. It is an open question as to how these theoretical works generalize to the case of partially coherent object distributions. 
\par
\section{Appendix}

\subsection{Derivation of Eqs. (\ref{eq:FI_Tsang},\ref{eq:FI_Object_plane})}
Assuming Poisson statistics and using Eqs. (\ref{eq:modal_weights}), which are derived in the supplement, we derive the FI matrix as
\begin{linenomath*}
\begin{align}
J&=\sum_{i=1,2}
\begin{bmatrix}
\frac{1}{N_{i}}(\frac{\partial N_{i}}{\partial \delta})^2 & \frac{1}{N_{i}}(\frac{\partial N_{i}}{\partial \delta})(\frac{\partial N_{i}}{\partial N_{0}})\\
\frac{1}{N_{i}}(\frac{\partial N_{i}}{\partial N_{0}})(\frac{\partial N_{i}}{\partial \delta}) & \frac{1}{N_{i}}(\frac{\partial N_{i}}{\partial N_{0}})^2
\end{bmatrix},\\
    &=
    \begin{bmatrix}
N_{0}\frac{\delta^2 d^2}{8\sigma^4}\left(\frac{1-\gamma d}{1-d^2}\right) & -\frac{\gamma\delta d}{2\sigma^2}\\
-\frac{\gamma\delta d}{2\sigma^2} & \frac{2}{N_{0}}(1+\gamma d)
\end{bmatrix}.
\end{align}
\end{linenomath*}
If only $\delta$ is unknown, the FI normalized by the object plane photon number $2N_{0}$ is described by $J(1,1)/2N_{0}$, which is equivalent to Eq. (\ref{eq:FI_Tsang}) and is plotted in Fig. (\ref{fig::1_partial_coherence_paper}a).\par
If both $\delta$ and $N_{0}$ are unknown, the FI for estimating $\delta$ is found by $\left(C\left(1,1\right)\right)^{-1}$, where $C=J^{-1}$, and $\left(C\left(1,1\right)\right)$ gives the multiparameter CRB. The matrix $C$ is given as
\begin{linenomath*}
\begin{align}
    C=\frac{1}{\frac{\delta^2d^2}{4\sigma^4}\left(\frac{1-\gamma^2}{1-d^2}\right)}
        \begin{bmatrix}
\frac{2}{N_{0}}(1+\gamma d) & \frac{\gamma\delta d}{2\sigma^2}\\
\frac{\gamma\delta d}{2\sigma^2} & N_{0}\frac{\delta^2 d^2}{8\sigma^4}\left(\frac{1-\gamma d}{1-d^2}\right) 
\end{bmatrix}.
\end{align}
\end{linenomath*}
The quantity $\left(C\left(1,1\right)\right)^{-1}/2N_{0}$ is given by Eq. (\ref{eq:FI_Object_plane}) and plotted in Fig. (\ref{fig::1_partial_coherence_paper}b).\par
\subsection{Experimental Details}
Please refer to the supplement for the experimental details about the field generation, the measurement of the modal weights, mode intensity versus $\delta$, and the data processing. The supplement is available under the `ancillary files' link on the arXiv page.
 \section*{Acknowledgements} The authors acknowledge Prof. J. R. Fienup, Dr. Walker Larson, Prof. Mankei Tsang, and Prof. Bahaa E. A. Saleh for useful discussions.
\section*{Funding Information}
 A.N.V, S.A.W, and K. L acknowledge support from the DARPA YFA \# D19AP00042. J. Y and A. N. J acknowledge support from NSF under award OMA-1936321. M. A. A. acknowledges support from the National Science Foundation (PHY-1507278) and the Excellence Initiative of Aix-Marseille University— A*MIDEX, a French ``Investissements d’Avenir'' program. R. W. B acknowledges support from the Office of Naval Research of the US (Award: N00014-19-1-2247)  and  the Natural Sciences and Engineering Research Council of Canada (RGPIN/2017-06880).



\end{document}